\documentclass[conference]{IEEEtran}
\IEEEoverridecommandlockouts
\usepackage{cite}
\usepackage{amsmath,amssymb,amsfonts}
\usepackage{algorithmic}
\usepackage{graphicx}
\usepackage{textcomp}
\usepackage{xcolor}
\def\BibTeX{{\rm B\kern-.05em{\sc i\kern-.025em b}\kern-.08em
    T\kern-.1667em\lower.7ex\hbox{E}\kern-.125emX}}
\usepackage{bbm}
\usepackage{multirow}
\usepackage{algorithm}
\newtheorem{theorem}{\textbf{Theorem}}
\begin{document}

\title{Adversarial for Social Privacy: A Poisoning Strategy to Degrade User Identity Linkage}

\author{\IEEEauthorblockN{Jiangli Shao,Yongqing Wang,Boshen Shi,Hao Gao,Huawei Shen,Xueqi Cheng}
\IEEEauthorblockA{\textit{Data Intelligence System Research Center, Institute of Computing Technology, Chinese Academy of Sciences}}
\{shaojiangli19z,wangyongqing,gaohao,shenhuawei,cxq\}@ict.ac.cn,527515645@qq.com
}

\maketitle

\begin{abstract}
Privacy issues on social networks have been extensively discussed in recent years. The user identity linkage (UIL) task, aiming at finding corresponding users across different social networks, would be a threat to privacy if unethically applied. The sensitive user information might be detected through connected identities. A promising and novel solution to this issue is to design an adversarial strategy to degrade the matching performance of UIL models. However, most existing adversarial attacks on graphs are designed for models working in a single network, while UIL is a cross-network learning task. Meanwhile, privacy protection against UIL works unilaterally in real-world scenarios, i.e., the service provider can only add perturbations to its own network to protect its users from being linked. To tackle these challenges, this paper proposes a novel adversarial attack strategy that poisons one target network to prevent its nodes from being linked to other networks by UIL algorithms. Specifically, we reformalize the UIL problem in the perspective of kernelized topology consistency and convert the attack objective to maximizing the structural changes within the target network before and after attacks. A novel graph kernel is then defined with Earth mover’s distance (EMD) on the edge-embedding space. In terms of efficiency, a fast attack strategy is proposed by greedy searching and replacing EMD with its lower bound. Results on three real-world datasets indicate that the proposed attacks can best fool a wide range of UIL models and reach a balance between attack effectiveness and imperceptibility.
\end{abstract}


\section{Introduction}
The privacy concern has attracted great attention on social networking services recently~\cite{privacy,pri2}. The sensitive user information may be unintentionally divulged by user identity linkage (UIL)~\cite{kong2013inferring,zhan2017community,trung2020comparative} which aims to link socialized user data across different social networks. Associated with cross-network linkages, user properties can be accurately aligned~\cite{shu2017user} and thus privacy data, including family, job, age, email, etc., can be easily revealed. Therefore, the lack of management on UIL would hold a candle to the devil. 
\begin{figure}[!ht]
	\centering
	\includegraphics[width=120pt,height=40pt]{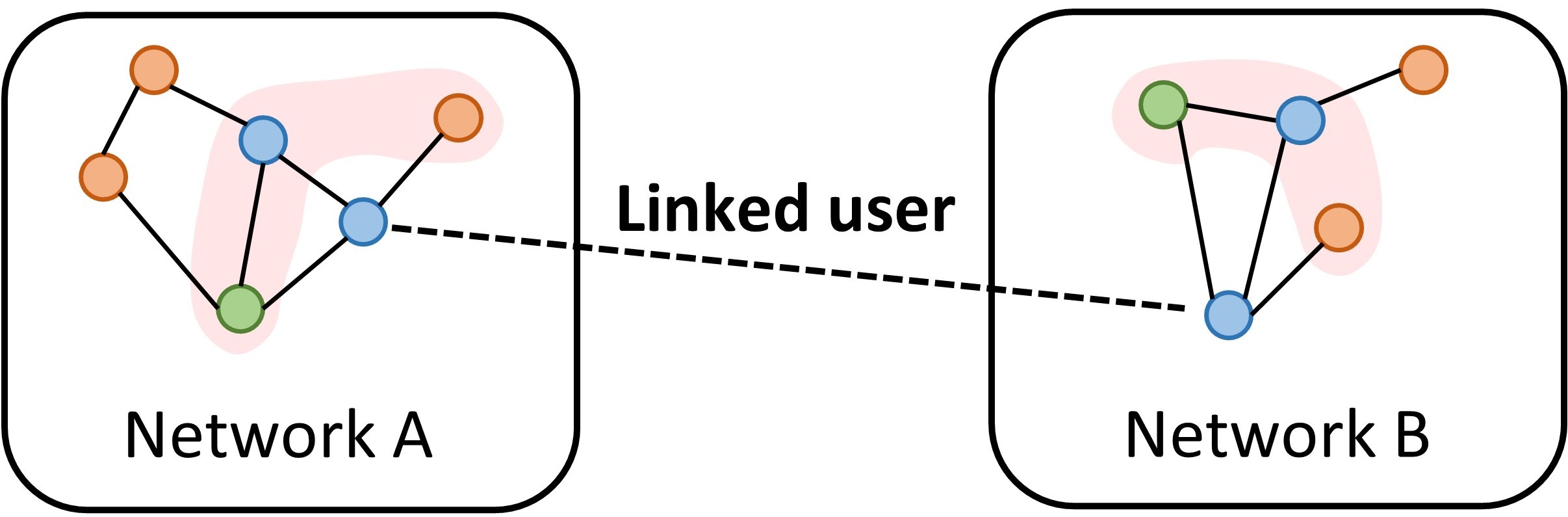}
	\caption{A sketch to display the topology consistency: the two nodes are supposed to be linked as they share similar neighborhoods.}
	\label{a}
\end{figure}

To combat the user privacy leakage problem led by malicious usage of UIL, the key is to degrade matching performance of UIL models when applied in practice. According to most literature, social relationships play a pivotal role in exploring corresponding user identities across networks~\cite{shu2017user,ione,man2016predict,deeplink,zhang2016final}. Typically, users who share similar neighborhoods in different graphs could be recognized as matched ones, which is known as topology consistency~\cite{zhang2016final,yan2021bright} (shown in Fig.~\ref{a}). Thus, disturbing topology consistency can lead to performance degradation on a series of UIL models relied on structures.

Previous studies~\cite{sun2018adversarial} on adversarial attacks have shown that imposing slight perturbations in graph structure can drastically worsen the performance of graph models, e.g., node classification~\cite{netattack,rls2v,metaself}, link prediction~\cite{aamas,fpta}, and graph classification~\cite{backdoor}. However, these adversarial attacks are constrained to fool learning models in a single network while UIL needs to consider multiple networks. To degrade matching performance across networks, GMA~\cite{gma} adopts a perturbation method calculated by node density estimation on two graphs, aiming to increase the possibility of deriving wrong matching. Although perturbations on two graphs take effects on degrading UIL models, it is still hard to be generalized into real applications. This is because the authority of perturbations for privacy protection can hardly be extended to multiple networks in practice.

To tackle these challenges, we propose a novel poisoning strategy named TOAK (\textbf{\underline{T}}opology-\textbf{\underline{O}}riented \textbf{\underline{A}}ttack via \textbf{\underline{K}}ernel method) to attack a wide range of UIL models via removing edges in a single network. The key idea of TOAK is to vastly break the topology consistency so that UIL models cannot identify similar users on different social networks. We first reformalize  UIL problem by graph kernels~\cite{yanardag2015deep,scholkopf2000kernel}, which measures similarity between two structural objects by kernel functions defined in reproducing kernel Hilbert space (RKHS). Based on kernel-based modeling methods, we prove an equivalent attack objective on minimizing kernel function values between clean and poisoned target network (i.e., maximizing the structural changes in the target network). A novel kernel based on earth mover’s distance (EMD)~\cite{rubner2000earth} in edge-embedding space is also proposed. Extensive experimental results on real datasets demonstrate the effectiveness of our proposed strategy in preventing user privacy leakage via UIL.

The main contributions of this paper are listed as follows:

\begin{itemize}
	\item{}\textbf{Problem formulation.}  We reformulate UIL problem on attribute graph with a unified Kernel-based Framework and a closed-form solution is finally obtained. It  provides new insight on how to efficiently break topology consistency so as to worsen performances of UIL models.
	
	\item{} \textbf{Novel poisoning strategy.} With the new perspective on UIL, we propose the TOAK method to attack common UIL models. TOAK converts the attack objective to minimize kernel function values within the clean and poisoned target network. Meanwhile, a novel graph kernel is proposed based on EMD~\cite{rubner2000earth} among edge embedding distributions that are obtained via variational graph autoencoder (VGAE)~\cite{kipf2016variational} and the random walk process.
	
	\item{} \textbf{Optimized computation.} We utilize a greedy strategy to remove edges to avoid repeated calculation. Furthermore, a lower bound is used to approximate EMD in the attacking objective, which reduces the complexity from $O(n^3log(n))$ to $O(n)$. The optimized method is more time-saving.
	
	\item{} \textbf{Performance.} We conduct extensive experiments to validate the performance of proposed methods. The experimental results demonstrate that (1) \textit{Transferability}. TOAK can be broadly applied to attack typical unsupervised and supervised UIL models; (2) \textit{Attack performance}. Guided by proposed attacks, the quality of UIL models shows a higher mismatching rate than the attacks from other state-of-the-art attack models. Meanwhile, the proposed attack model still achieves a good balance between effectiveness and imperceptibility than compared methods; (3) \textit{Efficiency}. The proposed acceleration strategy leads to more than 10$\times$ speed-up while receiving almost similar attack performance compared with the exact unaccelerated version.
	
\end{itemize}

\section{Related Works}
\subsection{Network Alignment}
Network alignment aims to search related nodes across different networks and has been widely applied in diverse research fields~\cite{shu2017user}, including chemistry~\cite{smalter2008gpm}, bioinformatics~\cite{singh2008global}, knowledge graph~\cite{shu2017user}, and social computing~\cite{kong2013inferring}. In this paper, attentions are paid to network alignment models in social fields. Sometimes, it is also known as user identity linkage~\cite{mu2016user} or anchor link prediction~\cite{kong2013inferring}.

Most previous works utilize the “topology consistency” in such kind of matching issues. Inspired by PageRank~\cite{page1999pagerank}, IsoRank~\cite{singh2008global} and FINAL~\cite{zhang2016final} algorithm iteratively propagate the pairwise node similarity in graph. Additionally, BigAlign~\cite{koutra2013big} uses the alternating projected gradient descent to solve the bipartite network alignment. Then, with the development of graph representation, topology consistency can be also addressed by specific embedding. PALE~\cite{man2016predict} and IONE~\cite{ione} learn cross-network mapping via embedding that captures the specific structural regularities. DeepLink~\cite{deeplink} uses embeddings inspired by word embedding technologies. Based on heterogeneous networks, TALP~\cite{li2020type} design a type-aware embedding method to better model matching objectives. 
\subsection{Adversarial Attack on Graph}
Generally, graph adversarial attack methods can be divided into three categories according to specific tasks, i.e., node-relevant, link-relevant, and graph-relevant tasks. Node-relevant models focus on node-level tasks such as node classification and node embedding. Z{\"u}gner et al.~\cite{netattack} first propose NETTACK to fool the node classification models. RL-S2V~\cite{rls2v} utilizes a reinforcement learning method to learn the attack policy. Additionally, Bojchevski et al.~\cite{dw3} analyze adversarial vulnerability by random walks and propose approaches to worsen the quality of node embeddings. Most link-relevant models attack link prediction tasks. Zhou et al.~\cite{aamas} try to minimize the similarity metrics to lower the accuracy of link prediction. FPTA~\cite{fpta} designs heuristic rules to attack target models. Moreover, graph-relevant models concentrate on graph-level tasks such as graph classification. Xi et al.~\cite{backdoor} propose the backdoor attack on graphs to worsen the model.

However, most attack models concentrate on single network tasks while UIL is related to multiple networks, thus existing models may be invalid in degrading the performance of UIL. Although the work like GMA~\cite{gma} can perform its effectiveness on graph matching, the requirement on cooperative perturbations cross networks limits the real application on privacy protection.

\section{Notations and Preliminaries}
This section introduces basic concepts used in the paper.

\subsection{Notations}
In this paper, a graph is defined as a tuple $G=(V,E,\boldsymbol{X})$, where $V$ and $E$ denote the set of nodes and edges respectively. Symbol $v$ denotes a node and $e_{ij}$ represents the edge with node $v_i$ and $v_j$ as its endpoints. Let $|V|$ and $|E|$ denote the total number of nodes and edges. Each node $v\in V$ has $d$-dimensional attributes which are represented by a row in $\boldsymbol{X}\in\mathbb{R}^{|V|\times d}$.  The adjacency matrix of $G$ is represented by $\boldsymbol{A}$. 

For a node $v\in V$, the $k$-hop neighborhood of $v$ contains nodes at a distance less than or equal to $k$ from $v$. The $k$-ego network of node $v$ is a subgraph of $G$ induced by the $k$-hop neighborhood of $v$ and $v$ itself, and is denoted by $G_{k}(v)$.

\subsection{Graph Kernel}
\label{sec:graph_kernel}
Kernel methods are commonly used to measure similarity between two objects with a kernel function which corresponds to an inner product in reproducing kernel Hilbert space (RKHS). Kernels on graphs are generally defined on R-Convolution framework~\cite{haussler1999convolution}. The main idea of R-Convolution is to decompose a graph into sub-structures and kernel value is a combination of sub-structure similarities. In this way, the kernel $\mathcal{K}(\cdot, \cdot):\mathbb{G}\times \mathbb{G}\rightarrow \mathbb{R}$ defined on graphs $G,G^*\in \mathbb{G}$ is given by:
\begin{align*}
	\mathcal{K}(G,G^*) = \sum_{g\in \mathcal{R}^{-1}(G)} \sum_{g^*\in \mathcal{R}^{-1}(G^*)} \prod_{d=1}^{D}\mathcal{K}_d(g_d, g^*_d),
\end{align*}
where $g=g_1,\ldots,g_D$ are decompositions of graph $G$, $\mathcal{R}$ denotes the relation from sub-structures $g$ to graph $G$, and $\mathcal{R}^{-1}(G)$ is a set of sub-structures in graph $G$. The $\mathcal{K}_d(\cdot,\cdot):\mathbb{R}^n\times \mathbb{R}^n\rightarrow \mathbb{R}$ is a general kernel and satisfies:
\begin{align*}
	\mathcal{K}_d(\boldsymbol{x}, \boldsymbol{x}^{\prime})=<\phi(\boldsymbol{x}),\phi(\boldsymbol{x}^{\prime})>.
\end{align*}
Here, function $\phi$ maps a vector to a high dimensional space, and $<\cdot,\cdot>$ denotes the inner product.

\subsection{Earth Mover's Distance}
Let $\mathcal{X}$ be a $d$-dimensional compact set and $\text{Prob}(\mathcal{X})$ denotes space of probability measures defined on $\mathcal{X}$. Earth mover's distance (EMD) defines the discrepancy between two distribution $P_{1}, P_{2}\in \text{Prob}(\mathcal{X})$:
\begin{align*}
	EMD(P_{1}, P_{2}) & = \inf_{\gamma\in \Gamma(P_{1}, P_{2})} \int\int_{n_1\times n_2} c(x,y)d\gamma(x,y),
\end{align*}
where $\Gamma(P_{1}, P_{2})$ is the set of all joint distributions whose marginals are $P_{1}$ and $P_{2}$. The $\gamma(x, y)$ can be intuitively interpreted as ``mass'' transportation from $x$ to $y$ so that EMD is the minimum cost for transforming one distribution $P_1$ into another distribution $P_2$. The $c(x,y)$ determines transportation cost from $x$ to $y$, which is generally derived from distance measures. The discrete form of EMD is as follows:
\begin{align*}
	EMD(P_{1}, P_{2}) = \inf_{\gamma\in \Gamma(P_{1}, P_{2})} \sum_{i=1}^{n_1}\sum_{j=1}^{n_2} c(x_i,y_j) \gamma(x_i,y_j).
\end{align*}

\subsection{Variational Graph Autoencoder}
Variational graph auto-encoder (VGAE)~\cite{kipf2016variational} is a commonly used network embedding framework, which encodes both graph structures and node attributes into the embedding. Given adjacency matrix $\boldsymbol{A}$ and attribute $\boldsymbol{X}$ for a graph, the encoder part of VGAE computes latent variables $\boldsymbol{\mu}$ and $log\boldsymbol{\sigma}$ with GCNs~\cite{gcn}: $\boldsymbol{\mu}=GCN_{\mu}(\boldsymbol{A},\boldsymbol{X})$, $ log\boldsymbol{\sigma}=GCN_{\sigma}(\boldsymbol{A},\boldsymbol{X})$, and latent embedding $\boldsymbol{z_i }$ is given by a normal distribution $\mathcal{N}$:
\begin{align*}
	q(\boldsymbol{z}_i\vert \boldsymbol{A},\boldsymbol{X})=\mathcal{N}(\boldsymbol{z}_i\vert \mu_i,diag(\sigma_i^2)).
\end{align*}
The decoder part reconstructs the graph via inner product of latent embeddings. In this paper, we use VGAE to generate node embeddings that combine structural and attribute features.

\section{Attacks on UIL Models}
\subsection{Problem Formulation}
Given source network $G^{s}=(V^{s},E^{s},\boldsymbol{X}^{s})$ and target network $G^{t}=(V^{t},E^{t},\boldsymbol{X}^{t})$. Let $\boldsymbol{H}\in [0,1]^{|V^{s}|\times |V^{t}|}$ represents the prior knowledge matrix.  The goal of UIL algorithm is to find the matching matrix:
\begin{displaymath}
	\boldsymbol{M} = UIL(G^{s},G^{t},\boldsymbol{H}).
\end{displaymath}
The matrix $\boldsymbol{M}$ has size $|V^{s}|\times |V^{t}|$ and the entry $\boldsymbol{M}_{ij}$ reflect the probability that node $v_{i}$ (in source networ) and node $v_{j}$ (in target network) are linked. Typically, matrix $\boldsymbol{H}$ denotes labeled data for supervised model. Otherwise, $\boldsymbol{H}$ can also be pairwise node similarity. If no prior knowledge is known, all entries of $\boldsymbol{H}$ are set to 0.

In this paper, the goal of adversarial attacks on UIL algorithms is to remove a set of edges $E^{-}\subset E^{t}$ to maximize changes in matching matrix. Let $\boldsymbol{M^{*}}$ denotes the matching matrix between source network and poisoned target network $G^{t*}=(V^{t},E^{t}-E^{-},\boldsymbol{X}^{t})$, the objective is:
\begin{equation}
	\begin{aligned}
		&\max\ \parallel\boldsymbol{M}- \boldsymbol{M^{*}}\parallel_{F}\quad
		s.t.~~|E^{-}|\leqslant K, \notag
	\end{aligned}
\end{equation}
where $K$ is an allowed threshold of removed edges. Here, we only consider the removing edge attack because graphs in real applications are usually sparse, thus the solution spaces are relatively small compared with other kinds of attack. Meanwhile, removing edge is easy to implement on social platforms.

\subsection{Unified Kernel-based Framework on UIL}
Previous studies on UIL have reached a consensus~\cite{zhang2016final,yan2021bright} that, among different networks, nodes that have similar local structures are more likely to be linked, which is known as the ``topology consistency". Accordingly, instead of assuming one specific UIL algorithm as the threat model, we propose a unified framework to re-formalize the UIL problem based on the widely used topology consistency, which is conducive to devising an effective and universal strategy to degrade diverse UIL models. 

The local structure of a node $v$ can be well captured by the $k$-ego network of $v$ and the similarity across different structures is measured by graph kernels. Thus, consistency between node $v_i$ in graph $G^{s}$ and node $v_j$ in graph $G^{t}$ can be measured as: $\mathcal{K}(G_{k}^{s}(v_i),G_{k}^{t}(v_j))$. The UIL task can be uniformly expressed as the following form:
\begin{align}
	\label{def}
	\boldsymbol{M}=\underset{\boldsymbol{M}}{\max}\ \sum_{i=1}^{|V^{s}|}\sum_{j=1}^{|V^{t}|}\boldsymbol{M}_{ij}\mathcal{K}(G_{k}^{s}(v_i), G_{k}^{t}(v_j))-\alpha\parallel \boldsymbol{M}-\boldsymbol{H}\parallel.
\end{align}

The above equation means nodes with higher topology consistency should have a higher probability to be aligned and, at the same time, the final alignment matrix $\boldsymbol{M}$ needs to be consistent with prior knowledge matrix $\boldsymbol{H}$. Here, $\alpha$ is a hyperparameter that controls the importance of prior knowledge.

Let $\boldsymbol{\Phi}$ denote the Kernel Matrix and $\boldsymbol{\Phi}_{ij} = \mathcal{K}(G_{k}^{s}(v_i),G_{k}^{t}(v_j))$. Eq~(\ref{def}) can be written in the matrix form as follows:
\begin{align*}
	\boldsymbol{M}=\underset{\boldsymbol{M}}{\max}\ tr(\boldsymbol{M}\boldsymbol{\Phi^{T}})-\alpha\parallel \boldsymbol{M}-\boldsymbol{H}\parallel,
\end{align*}

The closed-form solution is obtained by setting its derivative to be zero:
\begin{align*}
	\dfrac{\partial\ [tr(\boldsymbol{M}\boldsymbol{\Phi^{T}})-\alpha\parallel \boldsymbol{M}-\boldsymbol{H}\parallel]}{\partial\ \boldsymbol{M}}=\boldsymbol{\Phi}-2\alpha(\boldsymbol{M}-\boldsymbol{H})=0,
\end{align*}
and finally we can get
\begin{equation}
	\boldsymbol{M}=\dfrac{1}{2\alpha}\boldsymbol{\Phi}+\boldsymbol{H}.
	\label{solution}
\end{equation}
Eq.~(\ref{solution}) gives a unified view of UIL tasks based on the topology consistency. With such a perspective, the matching matrix is a linear combination of the kernel matrix and the prior knowledge matrix. Although most existing UIL models do not use kernel-based modeling explicitly, the pairwise similarity comparison in them is closely related to kernel methods. For example, the matching results of FINAL~\cite{zhang2016final} are relevant to random walk kernels and the similarity comparison in DeepLink~\cite{deeplink} can be viewed as a variant of path-based kernels. 

Next, we will display our attack strategy based on the proposed unified kernel-based framework on UIL.

\subsection{Topology-oriented Attack via Kernel Method (TOAK)}
With the unified kernel-based framework on UIL, the matching matrix $\boldsymbol{M^{*}}$ between source network $G^{s}$ and poisoned target network $G^{t*}$ is calculated as:
\begin{equation}
	\boldsymbol{M^{*}}=\dfrac{1}{2\alpha}\boldsymbol{\Phi}^{*}+\boldsymbol{H}.
	\label{solution2}
\end{equation}
where $\boldsymbol{\Phi}^{*}$ denote the Kernel Matrix for source network and poisoned target network so that $\boldsymbol{\Phi}^{*}_{ij} = \mathcal{K}(G_{k}^{s}(v_i),G_{k}^{t*}(v_j))$. Integrating Eq~(\ref{solution}) and Eq~(\ref{solution2}), the original objective of the attack problem can be transformed into the following form:
\begin{align}
	\label{target}
	&\max\ \parallel\boldsymbol{M}- \boldsymbol{M^{*}}\parallel_{F} 	
	= \max\ \parallel\boldsymbol{\Phi}- \boldsymbol{\Phi^{*}}\parallel_{F}\notag\\
	&= \max\ \sum_{i=1}^{|V^{s}|}\sum_{j=1}^{|V^{t}|}(\mathcal{K}(G_{k}^{s}(v_i), G_{k}^{t}(v_j))-\mathcal{K}(G_{k}^{s}(v_i), G_{k}^{t*}(v_j)))^2.
\end{align}

Here, $k$-ego network $G_{k}^{t*}(v)$ changes as the removing edge set $E^{-}$ changes and the goal is to maximize the Eq~(\ref{target}). It can be noticed that the source network $G^{s}$ and the target network $G^{t}$ are both involved in the objective function which indicates the optimization process needs information about two networks. However, collecting details across different networks is usually difficult in real applications. Therefore, the following theorem is utilized to simplify the objective function:
\begin{theorem}
	For node $v_i$ in source network $G^{s}$ and node $v_j$ in target network $G^{t}$, the following relation holds:
	\begin{align*}
		&\max\ (\mathcal{K}(G_{k}^{s}(v_i), G_{k}^{t}(v_j))-\mathcal{K}(G_{k}^{s}(v_i), G_{k}^{t*}(v_j)))^2\notag\\
		& \Rightarrow \min\ \mathcal{K}(G_{k}^{t}(v_j),G_{k}^{t*}(v_j)).
	\end{align*}
	\label{t41}
\end{theorem}

The proof is referred to in Appendix~\ref{appendix_a}. Theorem~\ref{t41} indicates that maximizing the differences on kernel function values between source network and target network before and after the attack as in Eq.~(\ref{target}) is equivalent to minimizing the kernel values for the clean and poisoned target network. With this transformation, only the target network information is required and the objective is converted to:
\begin{align}
	\min\ \sum_{j=1}^{|V^{t}|} \mathcal{K}(G_{k}^{t}(v_j),G_{k}^{t*}(v_j))\quad
	s.t.\ |E^{-}|\leqslant K.
	\label{tg}
\end{align}


\subsection{Edge Distribution Distance (EDD) Kernels}
\begin{figure}[!htbp]
	\centering
	\includegraphics[width=\linewidth]{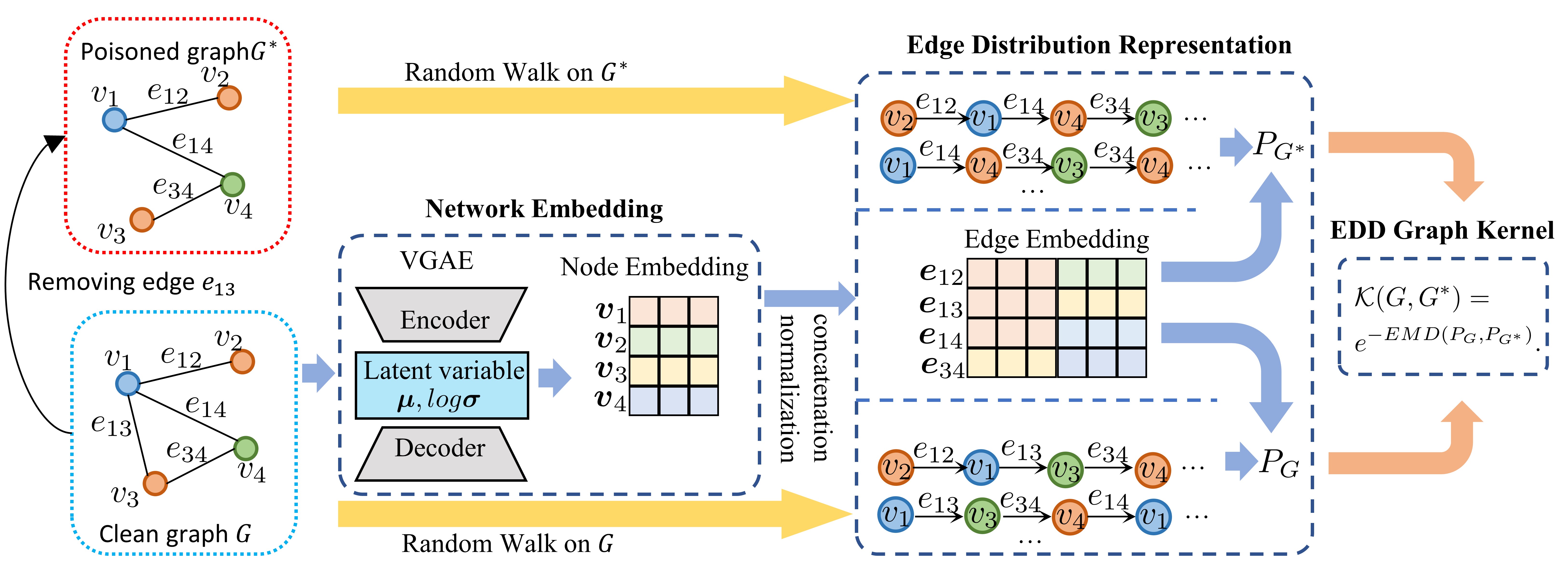}
	\caption{The model framework of proposed Edge Distribution Distance (EDD) kernel. The kernel first computes edge embeddings with VGAE. Then the graph is converted to the distribution on edge embedding space, where the probabilities of edges are given by the random walk. The final graph kernel is calculated based on earth mover's distance between two distributions.}
	\label{model}
\end{figure}
In this section, we will define a specific graph kernel applied in TOAK (shown in Fig.~\ref{model}). Following the R-Convolution framework mentioned in Section~\ref{sec:graph_kernel}, a naive decomposition treats the graph as a bag-of-edges, where pre-image $\mathcal{R}^{-1}(G)$ corresponds to the set of edges in graph $G$. In this way, the R-Convolution kernel related to $G$ and $G^*$ can be reduced to:
\begin{equation}
	\label{eq:basic_kernel}
	\mathcal{K}(G, G^*) = \prod_{d\in D} \mathcal{K}_d(\vec{e}_{ij}, \vec{e^*}_{ij}),
\end{equation}
where $D = \{(e_{ij}, e^*_{ij})|\ e_{ij}\in E,\ e^*_{ij}\in E^*\}$ refers to the set of comparable edge pairs between $G$ and $G^*$, and $\vec{e}_{ij},\ \vec{e^*}_{ij}$ denotes representations of edges $e_{ij},\ e^*_{ij}$ respectively.

Generally, edge representation can be well interpreted by its endpoints' representations~\cite{ione}. With consideration of the decomposition insensitive to topology variance in Eq.~(\ref{eq:basic_kernel}), we apply VGAE to generate powerful node representations that fuses the structural and attribute information: 
\begin{align*}
	& \boldsymbol{V_{emb}} = \text{VGAE}(\boldsymbol{A},\boldsymbol{X}),
\end{align*}
where $\boldsymbol{V_{emb}}$ is node representation matrix and row vector $\vec{v}_i$ denotes corresponding node representation, With the calculated node representation, the edge representation is obtained with concatenation and normalization of endpoints' representations:
\begin{align}
	\label{eemb}
	\vec{e}_{ij} = \dfrac{\vec{v}_{i}\oplus\vec{v}_{j}}{\parallel\vec{v}_{i}\oplus\vec{v}_{j}\parallel}.
\end{align}
Here, $\oplus$ represents the concatenation operation of two vectors and $\parallel\cdot\parallel$ denotes the norm of a vector.

Based on edge representations, kernel $\mathcal{K}_d$ is formalized as a radial basis function, i.e.,
\begin{equation}
	\label{eq:kernel_d}
	\mathcal{K}_d(\vec{e}_{ij}, \vec{e^*}_{ij})=\exp\ ({-\gamma_d\parallel\vec{e}_{ij}-\vec{e^*}_{ij}\parallel^2}),
\end{equation}
where $\gamma_d$ is a shape parameter to scale the distance between edges. Substitute Eq.~(\ref{eq:kernel_d}) into Eq.~(\ref{eq:basic_kernel}), the kernel $\mathcal{K}$ can be formulated as
\begin{equation}
	\label{eq:radial_kernel}
	\mathcal{K}(G, G^*) = \exp\ ({-\sum_{d\in D}\gamma_d\parallel\vec{e}_{ij}-\vec{e^*_{ij}}\parallel^2}).
\end{equation}

Then, we formalize the unresolved parameter $\gamma_d$. We regard a graph as a distribution on its edge embedding space and measures the discrepancy between two distributions with Earth mover's distance as follows:
\begin{equation}
	\begin{aligned}
		& EMD(P_G, P_{G^*}) =\hspace{-2mm} \inf_{\gamma\in \Gamma(P_{G}, P_{G^*})} \sum_{(e_{ij}, e^*_{ij})\in D}\hspace{-3mm} \gamma(e_{ij},e^*_{ij}) \|\vec{e}_{ij}-\vec{e^*_{ij}}\|^2 \\
		& \textit{s.t.~~} \sum_{e^*_{ij}} \gamma(e_{ij},e^*_{ij}) = p({e}_{ij}), \sum_{e_{ij}} \gamma(e_{ij},e^*_{ij}) = p(e^*_{ij}).
	\end{aligned}
	\label{gamad}
\end{equation}
Here $P_G=\{(\vec{e}_{ij},p(e_{ij}))\}_{e_{ij}\in E}$ and $P_{G^*}=\{(\vec{e^*}_{ij},p(e^*_{ij}))\}_{e^*_{ij}\in E^*}$ are the converted distribution of $G$ and $G^*$ respectively. Parameter $\gamma_d$ is exactly defined as the joint probability. With this definition, the kernel reflects not only the similarity between two graphs but also the importance of different edges in a graph. The effect of removing an edge can be calculated conveniently, which is essential in the attack task.

The marginal probability $p({e}_{ij})$ in Eq.~(\ref{gamad}) is defined as follows:
\begin{align}
	\label{prob}
	p({e}_{ij}) = \dfrac{RW({e}_{ij})\cdot\exp (\lambda\mathbb{I}(\boldsymbol{H}_{i:}+\boldsymbol{H}_{j:}))}{\sum_{{e}_{ij}} RW({e}_{ij})\cdot\exp (\lambda\mathbb{I}(\boldsymbol{H}_{i:}+\boldsymbol{H}_{j:}))},
\end{align}
where $RW(e)$ is a function counting the frequency that edge $e$ appears in random walk process. The $\boldsymbol{H}_{i:}$ and $\boldsymbol{H}_{j:}$ represent the sum of $i$-th row and $j$-th row in prior knowledge matrix $\boldsymbol{H}$ respectively. Function $\mathbb{I}(x)$ is an indicator, it equals 1 when $x\geqslant 1$ and equals 0 otherwise. Note that, random walk and prior knowledge indicate the edge importance in graphs and UIL task respectively.

Overall, the proposed kernel can be formulated as follows:
\begin{align}
	\label{emd}
	\mathcal{K}(G,G^{*})=e^{-EMD(P_{G}, P_{G^{*}})}.
\end{align}
The literature \cite{wwl}~and~\cite{geo} have proved kernels in this form are positive definite, thus ensuring the correctness of the problem-solving process in TOAK.

\section{Optimization and Analysis}
In this section, we equip the proposed kernel with a speed-up technique and then display the complete attack strategy. Some analyses on time complexity are given at the end.

\subsection{Acceleration}
Directly applying the proposed kernel to objective Eq.~(\ref{tg}) will calculate EMD with two $k$-ego networks. The time complexity of computing EMD in Eq.~(\ref{gamad}) is $O(n^3log(n))$, where $n$ represents the number of nodes in $k$-ego network. The high cost of computing makes it hard to be applied to dense or large-scale graphs. In this way, we estimate EMD by its lower bound:
\begin{theorem}
	EMD between $P_{G}$ and $P_{G^*}$ has the following lower bound:
	\begin{align*}
		EMD(P_{G},P_{G^*})
		\geqslant \bigg|\bigg|\sum_{e\in E}p(e)\vec{e}-\sum_{e^{*}\in E^{*}}p(e^{*})\vec{e^{*}}\bigg|\bigg|^{2}.
		\label{bound}
	\end{align*}
	\label{t51}
\end{theorem}
Proof and discussion on the lower bound can be found in Appendix~\ref{appendix_b}. Using the lower bound in Theorem~\ref{t51} to replace the EMD in Eq.~(\ref{emd}) and we have:
\begin{align}
	\mathcal{K}(G,G^{*})=\exp\ (-\bigg|\bigg|\sum_{e\in E}p(e)\vec{e}-\sum_{e^{*}\in E^{*}}p(e^{*})\vec{e^{*}}\bigg|\bigg|^{2}).
\end{align}

This form of kernel can be viewed as a variant of typical Gaussian kernels which is positive definite. Otherwise, the complexity is reduced to $O(n)$, which is applicable to large-scale datasets.
\subsection{Greedy Poisoning Strategy on Edges}
With the defined kernel functions, we can measure the impact of removing one edge within the object Eq~(\ref{tg}):
\begin{align}
	\label{score}
	score(e)&=\sum_{j=1}^{|V^{t}|} \mathcal{K}(G_{k}^{t}(v_j),G_{k}^{t*}(v_j)).
\end{align}
Here, poisoned graph $G^{t*}$ is obtained via removing edge $e$ in clean graph $G^{t}$. The graph $G_{k}^{t*}(v_j)$ is the $k$-ego network of node $v_j$ on poisoned graph. Note that, if edge $e$ is not in the edge set of $G_{k}^{t}(v_j)$, which means removing edge $e$ will not affect the $k$-ego network of $v_j$, thus the kernel value equals 1. Usually, $k$-ego network contains only a small amount of edges. Therefore, most terms in score(e) equal 1 and only a small part of kernels needs to be calculated.

After computing scores, the edge with the minimum score is removed. By repeating the computing-removing process for $K$ times, the set $E^{-}$ is finally obtained. However, recalculating the score for $K$ times is time-consuming. Therefore, we utilize the greedy selecting strategy. According to the score, we choose $K$ edges at once with $K$-smallest values to compose the removing edge set $E^{-}$. The overall algorithm is shown in Algorithm 1.

\begin{figure}[!ht]
\begin{algorithm}[H]
	\label{alg:algorithm}
	\caption{Topology Oriented Attack via Kernel method (TOAK)}
	\raggedright
	\textbf{Input}: The target network $G$, prior knowledge matrix $\boldsymbol{H}$.\\
	\textbf{Parameter}: Removing edge number $K$, coefficient $\lambda$. Random walk length $L$ and random walk amoumt $R$. \\
	\textbf{Output}: Removing edge set $E^{-}$.
	\vspace{-12pt}
	\begin{algorithmic}[1]
		\STATE Calculating the edge embedding as in Eq.~(\ref{eemb}).
		\STATE Conducting random walk with length $L$ on each node of $G$ for $R$ times and calculate the probability of edges in Eq.~(\ref{prob})
		
		\FOR{each edge $e$ in $G$}
		\STATE Calculate $score(e)$ as in Eq.~(\ref{score}).
		\ENDFOR
		
		\STATE Sorting $score(e)$ from small to large and selecting the first $K$ edges to compose $E^{-}$.
		\STATE \textbf{return} $E^{-}$.
	\end{algorithmic}
\end{algorithm}
\end{figure}

\subsection{Complexity Analysis}
The time complexity of the random walk process is $O(|V|RL)$, in which $R$ denotes random walk amount and $L$ denotes length of random walk. The score calculating has a complexity of $O(|E|nm)$, where the number $n$ represents the average nodes among the $k$-ego networks and the number $m$ denotes average times an edge appears in $k$-ego networks. The total complexity of TOAK is $O(|V|RL+|E|nm)$. As $|V|\gg R, |V|\gg L$ and $|E|\gg nm$, the algorithm is almost linear about node numbers and edge numbers. 

Note that the proposed score calculating method in Eq.~(\ref{score}) can also compute the effect of non-existing edges, which represents the adding-edge attack. However, due to the sparsity of graphs, the possible candidates of adding-edge attack are huge, causing nearly $O(|V|^2mn)$ time complexity of score calculating process. As a result, in this paper, we only consider removing-edges attacks. Efforts on designing other kinds of attacks will be our future task.

\section{Experiment}
\subsection{Experimental Setup}
In this section, we describe the datasets, baselines, UIL models, metrics, and other details of experiment\footnote{The code and data of TOAK will be available at Github after acceptance.}.

\textbf{Datasets} We conduct experiments on three real-world datasets to evaluate the performance of proposed TOAK model. Basic information of datasets is listed in Table~\ref{dataset}. The Douban and TF datasets come from famous social networks while the ARXIV dataset contains co-author networks. By reversing the source and target network in datasets, six user identity linkage datasets can be built and we denote them as: Douban (Online$\rightarrow$Offline), Douban (Offline$\rightarrow$ Online), TF (Twitter$\rightarrow$Foursqure), TF  (Foursquare$\rightarrow$Twitter), ARXIV (Mat$\rightarrow$DM), and ARXIV (CS$\rightarrow$DM),  where the former in brackets is the source network and the later is the target network. 
\begin{table}[!h]
	\caption{Dataset Description}
	\begin{tabular}{llccc}
		\hline
		Dataset & Network    & \#nodes & \#edges & \#linked users \\ \hline
		\multirow{2}{*}{Douban~\cite{db}}  & Online     & 3906   & 8164   & \multirow{2}{*}{1118}          \\
		& Offline    & 1118   & 1511   &               \\ \hline
		\multirow{2}{*}{TF~\cite{tf}}      & Twitter    & 5120   & 130575 & \multirow{2}{*}{3148}          \\
		& Foursquare & 5313   & 54233  &               \\ \hline
		\multirow{3}{*}{ARXIV~\cite{arxiv}}   & Mat        & 3506   & 7341   & \multirow{3}{*}{\shortstack{1400 (Mat-DM)\\3896 (CS-DM)}}  \\
		& DM         & 5465   & 14485  &    \\
		& CS         & 4946   & 11600  &    \\ \hline
	\end{tabular}
	\label{dataset}
\end{table}

\textbf{Baselines} Our TOAK model is compared with 6 baselines including state-of-the-art methods. Random attack randomly removes edges to generate perturbed graphs. DEG attack removes edges that connect to nodes with the largest degree and PR attack deletes edges connected to nodes with the highest pagerank~\cite{page1999pagerank} score. DW3~\cite{dw3} perturbs the random walk based node embedding via analysis on the PPMI matrix of the graph. FPTA~\cite{fpta} is a heuristic approach that relies on hand-crafted rules, which remove edges connected with known linked users. GMA~\cite{gma} push nodes to dense regions in graphs via the density estimation approach and achieves state-of-the-art performance. Also, we compare TOAK with its variants TOAK$_{-}$, which runs without the EMD acceleration technique. To ensure the fairness of comparison, baseline models only remove edges on the target network as TOAK does.

\begin{table}[!ht]
	\caption{Comparison among different target UIL methods.}
	\resizebox{\linewidth}{!}{
		\begin{tabular}{lcccc}
			\hline
			UIL method& Type & Gradient & Topology & Attribute \\ \hline
			FINAL~\cite{zhang2016final}    &unsupervised      & $\times$          &$\checkmark$          &$\checkmark$           \\
			REGAL~\cite{regal}    &unsupervised      &$\times$           &$\checkmark$          &$\checkmark$           \\
			PALE~\cite{man2016predict}     &supervised      &$\checkmark$           &$\checkmark$          &$\checkmark$           \\
			DeepLink~\cite{deeplink} &supervised      &$\checkmark$           &$\checkmark$          &$\times$           \\
			IONE~\cite{ione}     &supervised      &$\checkmark$           &$\checkmark$          &$\times$           \\ \hline
	\end{tabular}}
	\label{uilmodel}
\end{table}

\begin{table*}[!ht]
	\caption{Mismatching rate (\%) with removing 10\% edges on target network. Clean means the performance of UIL without any attacks. The $'-'$ means the attack methods cannot be applied to this UIL algorithm, thus have no reported results. Values with \textbf{boldface} and \underline{underline} refer to the best and second best results.}
	\resizebox{\linewidth}{!}{
		\begin{tabular}{l|ccccc|ccccc|ccccc}
			\hline
			& \multicolumn{5}{l|}{Douban (Online $\rightarrow$ Offline)}             & \multicolumn{5}{l|}{TF (Twitter $\rightarrow$ Foursquare)} & \multicolumn{5}{l}{ARXIV (Mat$\rightarrow$ DM)}              \\ \hline
			& FINAL & REGAL & PALE & DeePLink & IONE  & FINAL & REGAL & PALE & DeePLink & IONE & FINAL & REGAL & PALE & DeePLink & IONE   \\ \hline
			Clean       & 40.70          & 96.19          & 65.97          & 91.13          & 69.85          & 46.41          & 47.87          & 89.08          & 97.66          & 90.27          & 47.57          & 95.46          & 78.57          & 82.68          & 87.39          \\\hline
			Random      & 47.41          & 96.73          & 73.01          & 91.87          & 73.99          & 47.30          & 50.71          & 91.02          & 97.64          & 91.23          & 49.86          & 95.84          & 80.45          & 85.23          & 89.59          \\
			DEG         & 40.88          & 96.60          & 68.76          & 91.22          & 72.20          & 46.63          & 48.21          & 89.01          & 97.83          & 89.82          & 47.71          & 95.77          & 79.68          & 84.32          & 89.04          \\
			PR          & 42.75          & 96.92          & 68.80          & 91.75          & 72.74          & 47.20          & 48.24          & 89.15          & 97.78          & 89.88          & 47.57          & 95.86          & 80.79          & 83.55          & 90.43          \\
			DW3         & 42.93          & 96.80          & 73.45          & 92.54          & 75.42          & 46.57          & 51.55          & 92.05          & 98.09          & 92.12          & 45.43          & \underline{95.93}          & 80.54          & 83.80          & 89.14          \\
			FPTA        & 48.21          & 96.99          & 70.68          & 92.29          & 71.84          & 47.52          & 47.06          & 90.07          & 98.02          & 90.83          & 52.14          & 95.73          & 80.32          & 83.18          & 89.04          \\
			GMA         & -              & -              & 71.15          & 91.84          & 74.55          & -              & -              & 92.58          & 98.27          & 92.75          & -              & -              & 82.46          & 85.23          & 89.54          \\ \hline
			TOAK$_{-}$ & \textbf{55.01} & \underline{97.53}          & \textbf{76.29} & \underline{93.25} & \underline{76.31}          & \underline{50.07} & \underline{52.30}          & \underline{95.03}          & \textbf{99.38} & \textbf{95.89} & \underline{53.06}          & 95.89          & \underline{90.89}          & \underline{93.21}          & \underline{93.04}          \\
			TOAK         & \underline{54.95} & \textbf{97.78} & \underline{75.96}          & \textbf{93.27} & \textbf{76.52} & \textbf{50.13} & \textbf{52.69} & \textbf{95.89} & \underline{99.35} & \underline{95.35}          & \textbf{53.87} & \textbf{96.49} & \textbf{91.05} & \textbf{93.84} & \textbf{93.34} \\ \hline
	\end{tabular}}
	
	\resizebox{\linewidth}{!}{
		\begin{tabular}{l|ccccc|ccccc|ccccc}
			\hline
			& \multicolumn{5}{l|}{Douban (Offline $\rightarrow$ Online)}             & \multicolumn{5}{l|}{TF (Foursquare $\rightarrow$ Twitter)} & \multicolumn{5}{l}{ARXIV (CS$\rightarrow$ DM)}              \\ \hline
			& FINAL & REGAL & PALE & DeePLink & IONE  & FINAL & REGAL & PALE & DeePLink & IONE & FINAL & REGAL & PALE & DeePLink & IONE   \\ \hline
			Clean       & 41.50          & 96.15          & 64.49          & 85.14          & 70.68          & 46.41          & 44.10          & 85.03          & 96.94          & 90.36          & 15.48          & 90.36          & 67.80          & 71.72          & 84.39          \\\hline
			Random      & 46.60          & 95.87          & 69.90          & 85.79          & 72.67          & 46.54          & 47.33          & 86.28          & 97.44          & 91.77          & 18.35          & 93.56          & 70.05          & 75.38          & 86.23          \\
			DEG         & 43.29          & 96.12          & 65.50          & 84.76          & 72.40          & 46.95          & 45.65          & 85.91          & 97.26          & 91.00          & 16.45          & 90.66          & 69.12          & 72.94          & 85.72          \\
			PR          & 43.38          & 96.53          & 66.19          & 85.30          & 72.25          & 47.01          & 46.14          & 85.63          & 97.11          & 90.58          & 15.76          & 90.39          & 70.72          & 73.69          & 85.61          \\
			DW3         & 47.41          & 95.64          & 71.44          & 87.04          & 75.33          & 46.16          & 47.06          & 86.87          & 97.23          & 91.80          & 16.14          & 93.81          & 73.29          & 76.87          & 88.31          \\
			FPTA        & 47.50          & 96.91          & 71.22          & 88.27          & 72.78          & 46.47          & 45.59          & 85.88          & 97.65          & 91.46          & 20.74          & 93.25          & 69.36          & 76.59          & 86.38          \\
			GMA         & -           & -           & 69.50          & 87.26          & 72.36          & -           & -           & 89.51          & 98.78          & 93.63          & -           & -           & 73.44          & 77.83          & 85.08          \\ \hline
			TOAK$_{-}$ & \textbf{54.29} & \underline{97.01} & \underline{93.05}          & \textbf{96.34} & \textbf{89.03} & \underline{49.62} & \underline{51.08}          & \underline{93.26}          & \underline{99.01}          & \textbf{96.02} & \underline{26.17}          & \underline{93.92}          & \underline{80.98}          & \underline{87.95}          & \underline{90.07}          \\
			TOAK         & \underline{54.09}          & \textbf{97.09} & \textbf{93.72} & \underline{96.14}          & \underline{88.01}          & \textbf{49.71} & \textbf{51.97} & \textbf{94.06} & \textbf{99.67} & \underline{95.42}          & \textbf{27.10} & \textbf{94.62} & \textbf{81.99} & \textbf{88.63} & \textbf{90.30} \\ \hline	
	\end{tabular}}
	\label{res}
\end{table*}

\textbf{Target UIL models} The effectiveness of the proposed attack method is validated with kinds of UIL models, and differences among them are displayed in Table~\ref{uilmodel}. FINAL~\cite{zhang2016final} is a network alignment model which finds aligned nodes through iterations on the Kronecker product of adjacency matrix. REGAL~\cite{regal} leverages the power of automatically learned node representations to match nodes across different graphs. PALE~\cite{man2016predict} employs network embedding with awareness of observed anchor links as supervised information. DeepLink~\cite{deeplink} utilize more powerful embedding approaches inspired by word embedding technologies. IONE~\cite{ione} learns multiple node embeddings and tries to model followers/followees as different context vectors. Note that, FINAL and REGAL are not deep models, thus the GMA method can not be applied to them.

\textbf{Metrics} The accuracy index is the most popular measure in UIL tasks, which calculates the ratio of correctly linked user pairs among all linked pairs. To directly reflect the attacking quality, we use a variant of Accuracy named Mismatching Rate (MR) as the metric. MR can be calculated as follows:
\begin{align*}
	MR=1-Accuracy
\end{align*}

As for the measurement of imperceptibility of attacks, it is generally weighted by the total removed edges. However, this index can only reflect the imperceptibility of the whole graph. To go a further step, we consider the imperceptibility at a more precise node level, and computes the ratio of unpoisoned edges on each node, which is called Node Imperceptibility Score ($NIS$):
\begin{align*}
	NIS(v)=1-\dfrac{\big|\{(v,v_{i})|(v,v_{i})\in E^{-},\ v_i\in V \}\big|}{\big|\{(v,v_{i})|(v,v_{i})\in E,\ v_i\in V \}\big|}
\end{align*}
The higher the $NIS$, the harder the attack is to be perceived at the node $v$.

\textbf{Experimental Details}
In this part, we describe the implementation details of TOAK and the experimental settings.

TOAK utilizes typical VGAE\footnote{https://github.com/dmlc/dgl/tree/master/examples/pytorch/vgae} to embed the network. The latent variable of VGAE is calculated with 2-layers GCN with hidden dimensions of 32 and 16 respectively. The learning rate is set to 0.001 and the total running epochs are 1000. The prior knowledge matrix $\boldsymbol{H}$ in unsupervised UIL algorithms is the node degree similarity matrix and for supervised UIL models, $\boldsymbol{H}_{ij}=1$ for training node pairs $(v_i,v_j)$ and $\boldsymbol{H}_{ij}=0$ otherwise. For all of the datasets, the random walk length (hyperparameter $L$) is set to 5 and repeated 1000 times (hyperparameter $R$). The hyperparameter $\lambda=1.0$ in Douban (online $\rightarrow$ offline) and $\lambda=5.0$ for Douban (offline $\rightarrow$ online). For TF and ARXIV datasets, $\lambda$ are all set to $2.0$. 

As for the details of UIL algorithms. FINAL and REGAL are unsupervised algorithms, and all the linked users are used for testing. The supervised UIL algorithms including PALE, DeepLink, and IONE randomly select 20\% of linked user node pairs as training set and remains 80\% of linked user pairs for testing. The hyperparameters of UIL models are adjusted to the best. We repeat the experiment 5 times and report the average mismatching rate.

All of the experiments are conducted on a compute server running on Linux with 2 CPUs of Intel Xeon E5-2640 v4 (at 2.40 GHz), 2 GPUs of NVIDIA Telsa K80 (11GB memory), and 128GB of RAM.

\subsection{Overall Comparision}
Firstly, we make an overall comparison of the attack performance among all the models when removing 10\% edges in target network. The results are shown in Table~\ref{res}.

We can observe that the proposed model outperforms other baselines on all datasets. Among different experimental settings over datasets and UIL algorithms, the proposed TOAK improves mismatching rate with an average of 4.58\% when compared with best baselines, and gains at most 22.28\% increasing on PALE in Douban (Offline$\rightarrow$Online), indicating the effectiveness of TOAK. Meanwhile, results show that attacks from baselines perform unsteady over UIL models. Besides, with a constraint on gradient descent method, GMA cannot be applied to FINAL and REGAL. Meanwhile, TOAK has no such limits and gains the best performance, reflecting better transferability than compared attack methods.

It is worth noting that, on ARXIV datasets, barely removing edges in DM network can dramatically confused the linkage on both the Mat$\rightarrow$DM and CS$\rightarrow$DM. Such a result demonstrates the potential application of TOAK in real-world, where only the target network can be modified. Interestingly, TOAK always obtains more MR promotions on supervised deep UIL models. The phenomenon may be attributed to the vulnerability of deep linkage methods. Finally, despite approximated EMD used in TOAK, it can still achieve similar mismatching rate as TOAK$_{-}$ (which applies exact calculation on EMD). The comparison on TOAK and TOAK$_{-}$ demonstrates the correctness of the accelerating process.

\subsection{Ablation Study}
\begin{figure*}[!ht]
	\centering
	\includegraphics[width=0.85\linewidth]{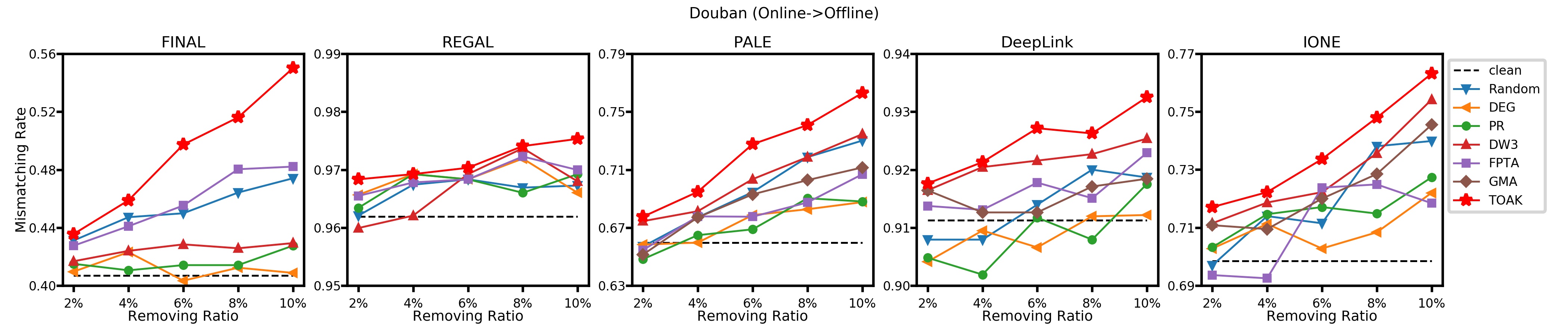}
	\includegraphics[width=0.85\linewidth]{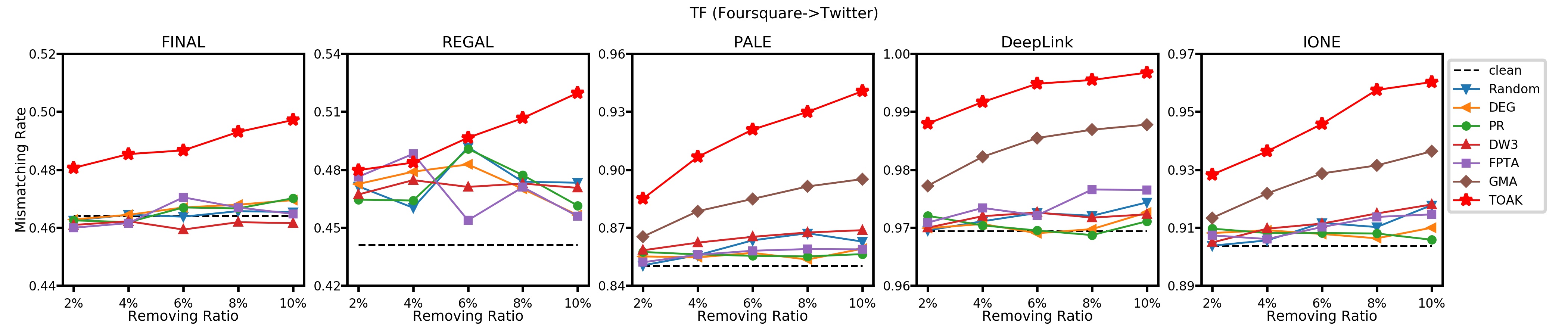}
	\caption{Mismatching rate on Douban (Online$\rightarrow$Offline) and TF (Foursquare$\rightarrow$Twitter) datasets when removing different ratios of edges in the target network.}
	\label{ratio1}
\end{figure*}
Fig.~\ref{ratio1} presents mismatching rate on various UIL models by varying the ratios of removed edges from 2\% to 10\%. As it illustrates, TOAK model always has higher mismatching rates under different ratios of removed edges when compared with other baselines. Meanwhile, mismatching rates can be consistently increased along with higher removing ratios resulting from TOAK.
\begin{figure}[!h]
	\centering
	\includegraphics[width=0.9\linewidth]{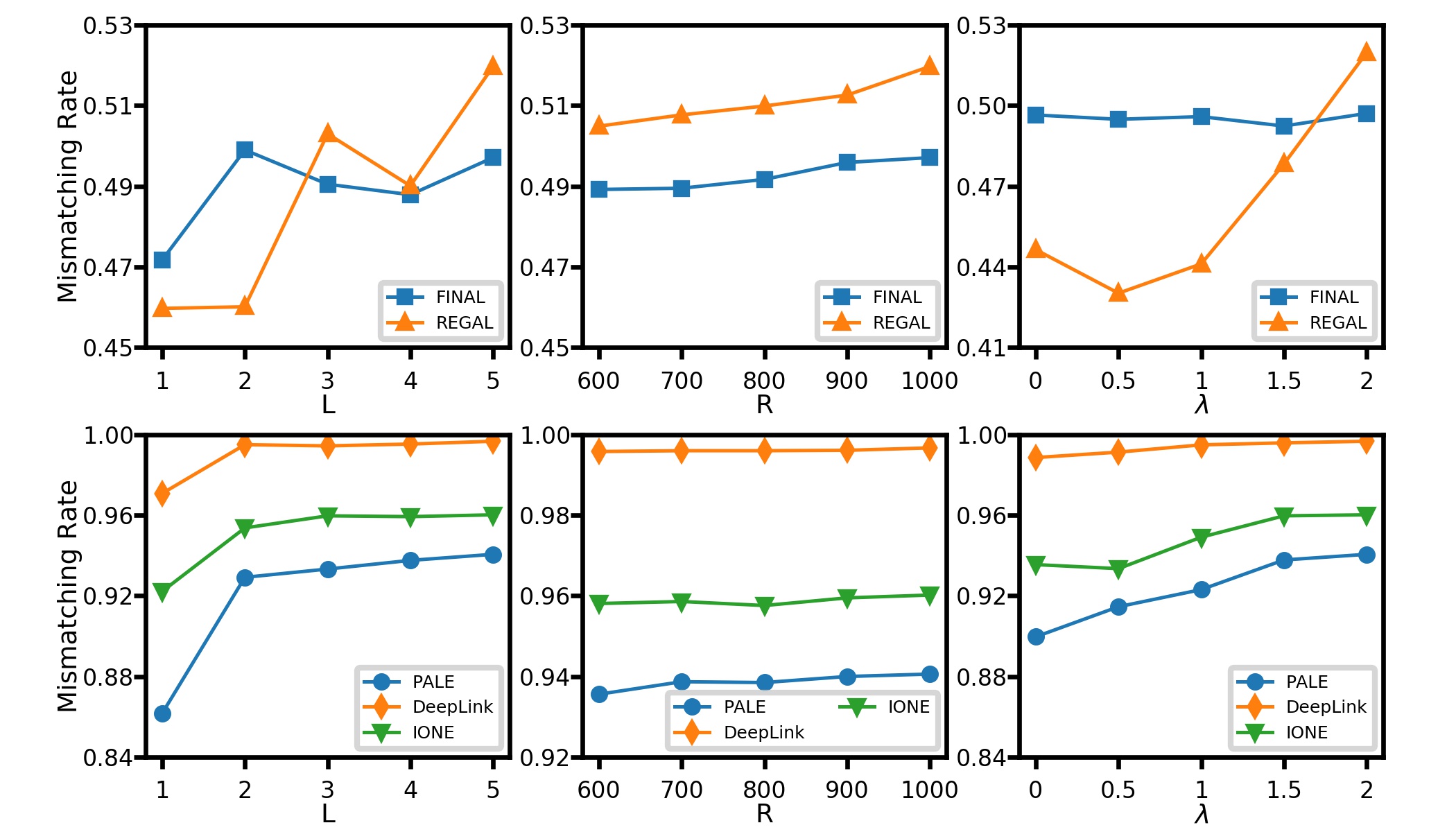}
	\caption{Effect of hyperparameters on the dataset TF (Foursquare $\rightarrow$ Twitter).}
	\label{hyper}
\end{figure}

Next, we discuss the influence of hyper-parameters in the proposed method, including random walk length $L$, random walk amount $R$, and coefficient $\lambda$. Comparisons are conducted with 10\% edges in the target network removed. The experimental results on TF (Foursquare$\rightarrow$Twitter) dataset are illustrated in Figure~\ref{hyper}. Besides, the influence on other datasets has similar trends and we show them in Appendix~\ref{appendix_c}. As for random walk length, we can notice that using a longer walk length can improve the performance, and the mismatching rate becomes stable when the length is extended. The phenomenon can be explained that the longer walk length can reflect more comprehensive local structures around nodes thus leading a better performance. In addition, random walk amount has no prominent influence on TOAK, where mismatching rate keeps similar as random walk amount varies from 600 to 1000. The last hyper-parameter $\lambda$ plays an important role in TOAK, which controls the ratio of the prior knowledge in edge distributions. TOAK has almost the worst performance when $\lambda=0$, which means no prior knowledge to refer to, and the mismatching rate starts rising as $\lambda$ becomes larger. Results on $\lambda$ indicate that prior knowledge plays an important role in attacking.

\begin{figure}[!ht]
	\centering
	\includegraphics[width=0.9\linewidth]{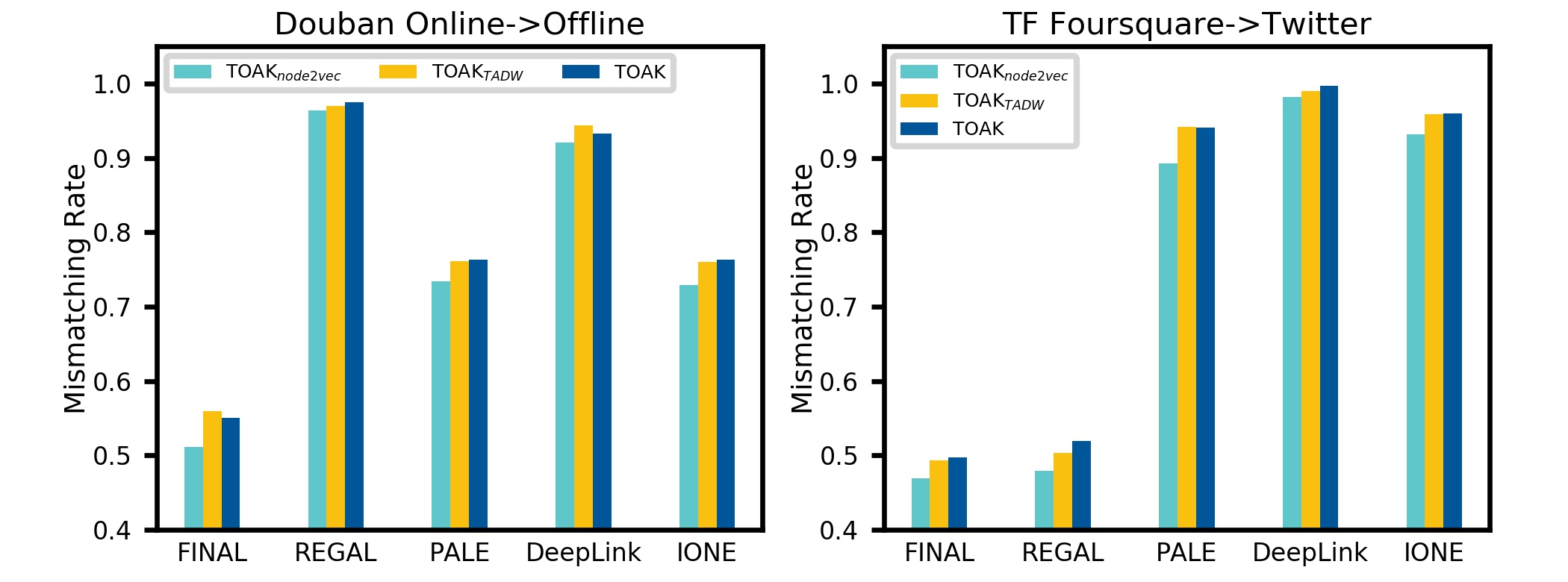}
	\caption{Performance comparison among different network embedding methods. (10\% edges in the target network are removed.)}
	\label{emb}
\end{figure}

Additionally, the effect of network embedding methods in TOAK is also verified and the results are shown in Fig.~\ref{emb}. Here TOAK is the proposed model with VGAE embedding. The TOAK$_{node2vec}$ and TOAK$_{TADW}$ uses node2vec~\cite{node2vec} and TADW~\cite{tadw} embedding respectively. The mismatching rate is similar on TOAK and TOAK$_{TADW}$ while TOAK$_{node2vec}$ has worse performance. It is because node2vec considers only the graph structure features and ignores the node attributes.

\subsection{Imperceptibility \& Efficiency}
\begin{figure}[!ht]
	\centering
	\includegraphics[width=0.9\linewidth]{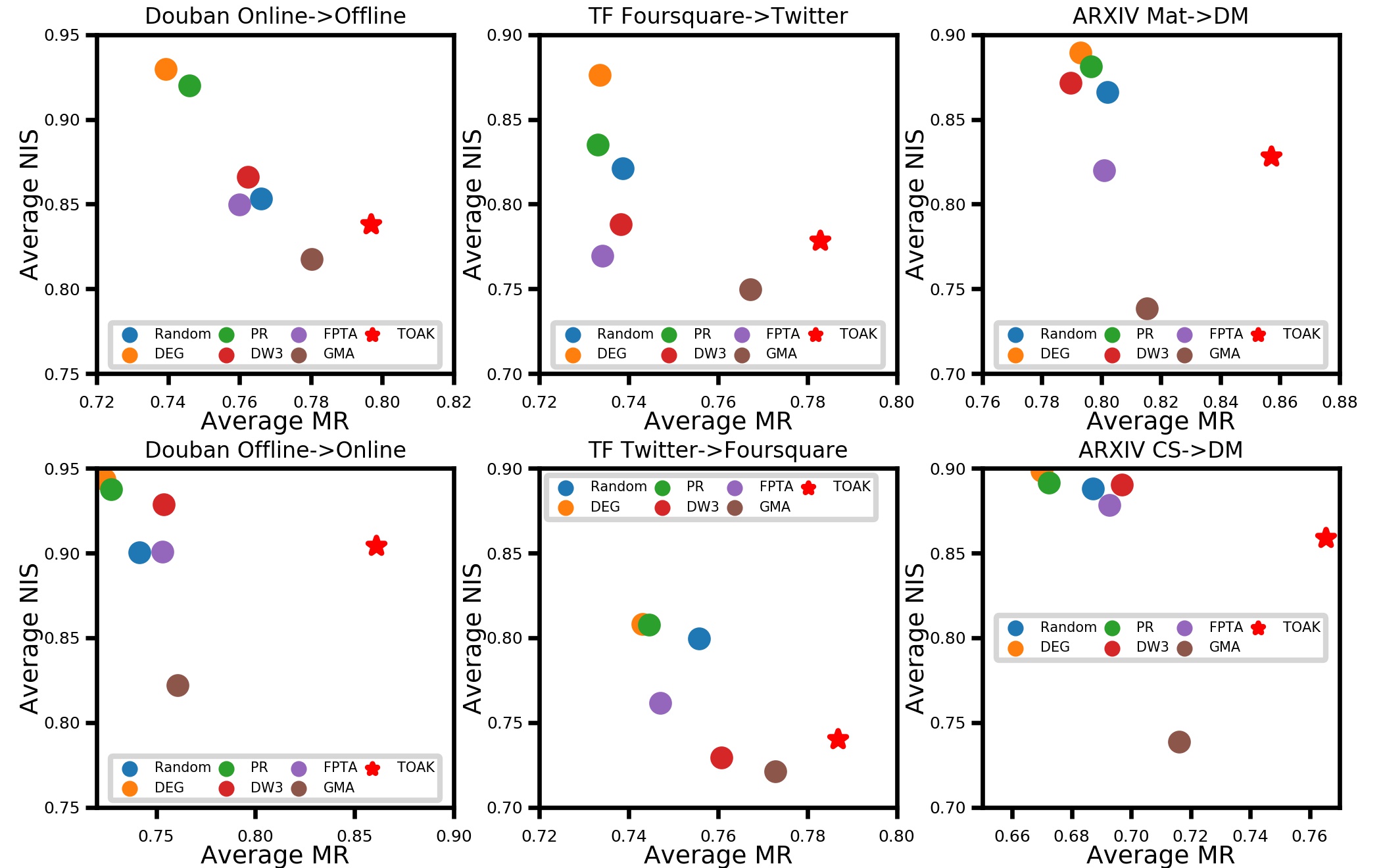}
	\caption{Balance between Mismatching rate and imperceptibility. (10\% edges in the target network are removed.)}
	\label{imp}
\end{figure}

In addition to the effectiveness of the model, the imperceptibility of attack and efficiency are also of great significance. Generally, the imperceptibility of the attack is weighted by the total removed edges, which can only reflect the attack imperceptibility on the whole graph. In practice, users are more sensitive to changes that appear closely around them. As a result, we consider the average node imperceptibility score to better reflect the attack imperceptibility of a single user. The results are shown in Fig~\ref{imp}. It is obvious that our model has the best mismatching rate performance and has higher imperceptibility than the state-of-the-art baseline GMA. Another baseline like DEG or PR has higher $NIS$ but their performances are limited. Overall, the proposed TOAK model can reach a balance between attack effectiveness and attack imperceptibility.

\begin{figure}[!ht]
	\centering
	\includegraphics[width=0.9\linewidth]{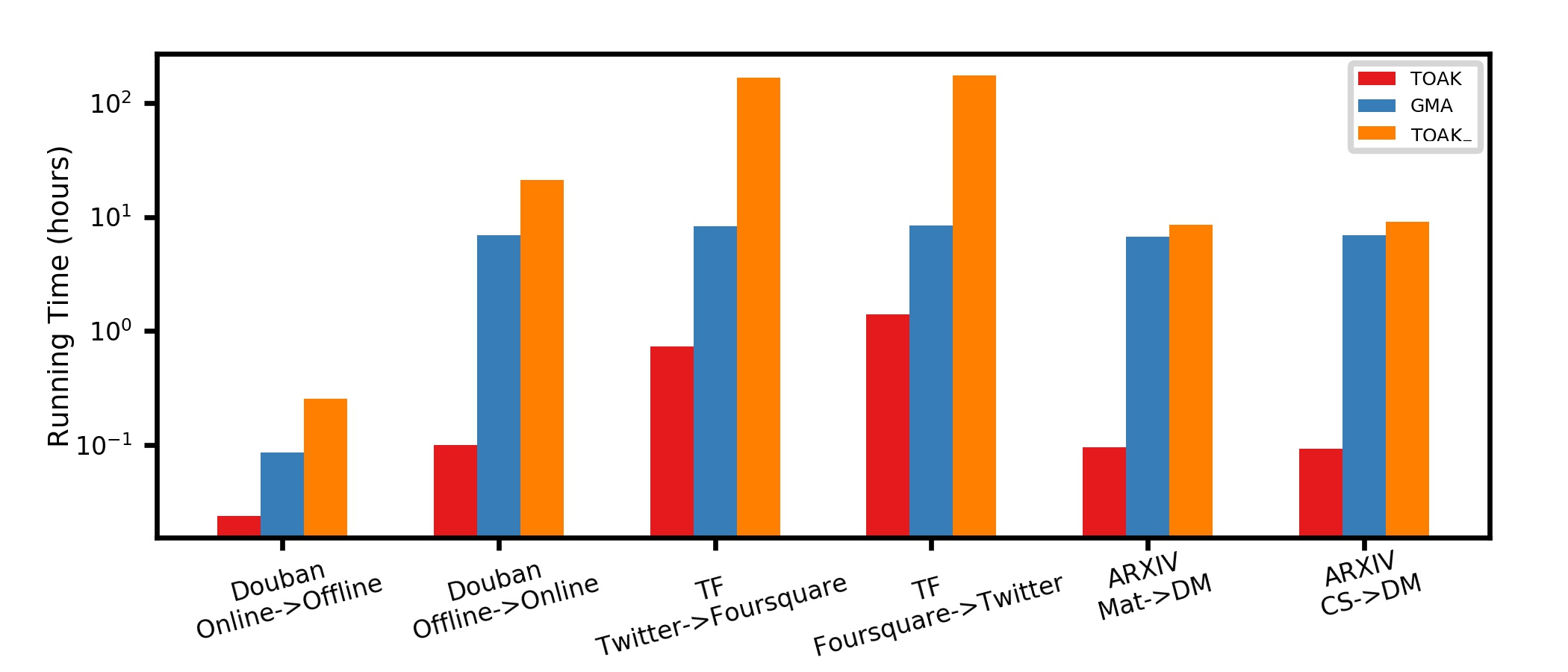}
	\caption{Runinng time on different datasets. (10\% edges in the target network are removed.)}
	\label{time}
\end{figure}

Fig~\ref{time} shows the time cost of GMA, TOAK, and TOAK$_{-}$ on different datasets for removing 10\% edges in the target network. The TOAK$_{-}$, which computes earth mover’s distance directly, spends much more running time than TOAK and GMA, especially for large target networks. With acceleration techniques, TOAK shows more than 10 times speedup compared with TOAK$_{-}$ among different datasets. The acceleration process becomes necessary when the graph is large. The TOAK and TOAK$_{-}$ have almost the same mismatching rate whereas TOAK has a lower time cost. At the same time, TOAK performs better than GMA and is less time-consuming. Comparison of running times shows the proposed TOAK is superior on not only the effectiveness but also the efficiency.

\section{Conclusion}
In this paper, we notice the potential privacy leakage caused by user identity linkage and design a novel strategy to degrade UIL algorithms via removing edges in a network. The proposed method re-formalizes the genera UIL problem in the graph kernel framework, which models the topology consistency. The attack problem is then converted to maximizing the structural changes in a single network. A novel kernel based on earth mover's distance among edge embedding distributions is proposed as well, and the attack strategy is designed according to the kernel. To make the proposed model more time-efficient, we adopt an approximate method and greatly decreased the complexity. Experimental results on real-world datasets show that TOAK achieves state-of-the-art performance and obtains a balance between attack effectiveness and imperceptibility.

In the future, we will consider more types of attacks, such as flipping edges or modifying node attributes, based on the proposed kernelized framework. Moreover, some heuristic strategies would be also introduced to improve the capability in more practical scenarios, e.g., partial observation of datasets. Also, the concealment of attacks should be meticulously considered. It might be conducive to explicitly model the imperceptibility in the objective function.

\bibliographystyle{IEEEtran}
\bibliography{IEEEabrv,myref1}

\appendix
\section{APPENDIX}
\subsection{Proof of Theorem~\ref{t41}}\label{appendix_a}
According to the definition of graph kernels, for any graph $G_1,G_2\in\mathbb{G}$, there exists a mapping function $\varphi:\mathbb{G}\rightarrow\mathbb{R}^n$ that satisfies:
$\mathcal{K}(G_1,G_2)=<\varphi(G_1),\varphi(G_2)>$,
where $<\cdot,\cdot>$ denotes the inner product. 

Therefore, for node $v_i$ in source network $G^{s}$ and node $v_j$ in target network $G^{t}$, we have:
\begin{align*}
	&(\mathcal{K}(G_{k}^{s}(v_i), G_{k}^{t}(v_j))-\mathcal{K}(G_{k}^{s}(v_i), G_{k}^{t*}(v_j)))^2\notag\\
	&\hspace{-9pt}=<\varphi(G_{k}^{s}(v_i),\varphi(G_{k}^{t}(v_j))-\varphi(G_{k}^{t*}(v_j))>^2\notag\\
	&\hspace{-9pt}=\hspace{-2pt}\varphi^2(G_{k}^{s}(v_i)[\varphi^2(G_{k}^{t}(v_j))\hspace{-2pt}+\hspace{-2pt}\varphi^2(G_{k}^{t*}(v_j))\hspace{-2pt}-\hspace{-2pt}2\varphi(G_{k}^{t}(v_j))\varphi(G_{k}^{t*}(v_j))]
\end{align*}

In the above equation, terms $\varphi^2(G_{k}^{s}(v_i)$, $\varphi^2(G_{k}^{t}(v_j))$ and $\varphi^2(G_{k}^{t*}(v_j))$ are the inner product of theirself. Thus they can be regarded as a constant. By ignoring the constant, we have:
\begin{align*}
	&(\mathcal{K}(G_{k}^{s}(v_i), G_{k}^{t}(v_j))-\mathcal{K}(G_{k}^{s}(v_i), G_{k}^{t*}(v_j)))^2\notag\\
	&\propto-\varphi(G_{k}^{t}(v_j))\varphi(G_{k}^{t*}(v_j))\notag
	=-\mathcal{K}(G_{k}^{t}(v_j),G_{k}^{t*}(v_j))
\end{align*}
The above equation proves that:
\begin{align*}
	&\max\ (\mathcal{K}(G_{k}^{s}(v_i), G_{k}^{t}(v_j))-\mathcal{K}(G_{k}^{s}(v_i), G_{k}^{t*}(v_j)))^2\notag\\
	&\Rightarrow \min\ \mathcal{K}(G_{k}^{t}(v_j),G_{k}^{t*}(v_j))
\end{align*}

\subsection{Proof of Theorem~\ref{t51}}\label{appendix_b}
Let $\gamma_{m}$ denotes the joint distribution that minimize the transportation cost from distribution $P_{G}$ to distribution $P_{G^*}$. According to the definition of EMD, we have
\begin{align}
	\label{ieq}
	&EMD(P_{G},P_{G^*}) =\sum_{(e_{ij}, e^*_{ij})\in D} \gamma_{m}(e_{ij},e^*_{ij}) \|\vec{e}_{ij}-\vec{e^*}_{ij}\|^2\notag\\
	\geqslant&\bigg|\bigg|\sum_{(e_{ij}, e^*_{ij})\in D} \gamma_{m}(e_{ij},e^*_{ij}) (\vec{e}_{ij}-\vec{e^*}_{ij})\bigg|\bigg|^2\\
	=&\bigg|\bigg|\sum_{(e_{ij}, e^*_{ij})\in D} \gamma_{m}(e_{ij},e^*_{ij}) \vec{e}_{ij}-\sum_{(e_{ij}, e^*_{ij})\in D}\gamma_{m}(e_{ij},e^*_{ij})\vec{e^*}_{ij}\bigg|\bigg|^2\notag\\
	=&\bigg|\bigg|\sum_{e_{ij}}\sum_{e^*_{ij}} \gamma_{m}(e_{ij},e^*_{ij}) \vec{e}_{ij}-\sum_{e_{ij}^*}\sum_{ e_{ij}}\gamma_{m}(e_{ij},e^*_{ij})\vec{e^*}_{ij}\bigg|\bigg|^2\label{lower}	
\end{align}
The inequality~(\ref{ieq}) is obtained via Jensen Inequality. Recall that, for any $\gamma\in \Gamma(P_{G}, P_{G^*})$:
\begin{align}
	\sum_{e^*_{ij}} \gamma(e_{ij},e^*_{ij}) = p({e}_{ij}),\quad \sum_{e_{ij}} \gamma(e_{ij},e^*_{ij}) = p(e^*_{ij}).\label{p}
\end{align}
With~(\ref{lower}) and~(\ref{p}), we have:
\begin{align*}
	EMD(P_{G},P_{G^*}) \geqslant&\bigg|\bigg|\sum_{e_{ij}}p(e_{ij}) \vec{e}_{ij}-\sum_{e_{ij}^*}p(e_{ij}^{*})\vec{e^*}_{ij}\bigg|\bigg|^2\\
	=&\bigg|\bigg|\sum_{e\in E}p(e)\vec{e}-\sum_{e^{*}\in E^{*}}p(e^{*})\vec{e^{*}}\bigg|\bigg|^{2}
\end{align*}
where $E$ and $E^*$ denote the edge set of $G$ and $G^{*}$ respectively.

\begin{figure}[!htbp]
	\centering
	\includegraphics[width=0.85\linewidth]{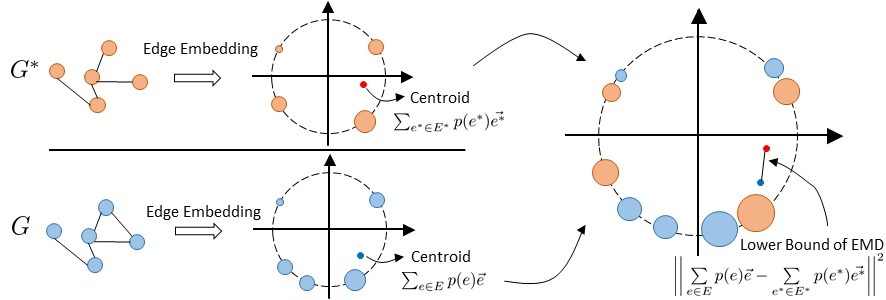}
	\caption{An explanation of the lower bound on EMD. Colored points are the normalized edge embeddings and the size of the point reflects the probability of an edge. The lower bound is exactly the centroid distance of two distributions. }
	\label{emdpic}
\end{figure}

An intuitive expression on this lower bound is shown in Fig~\ref{emdpic}. With the normalized edge embeddings, edges in a graph distributing on the unit circle. The centroid of a distribution is the probability-weighted sum of all edge embeddings in a graph. The lower bound of EMD calculates the distance between two centroids of distributions.

\begin{figure}[!ht]
	\centering
	\includegraphics[width=0.85\linewidth]{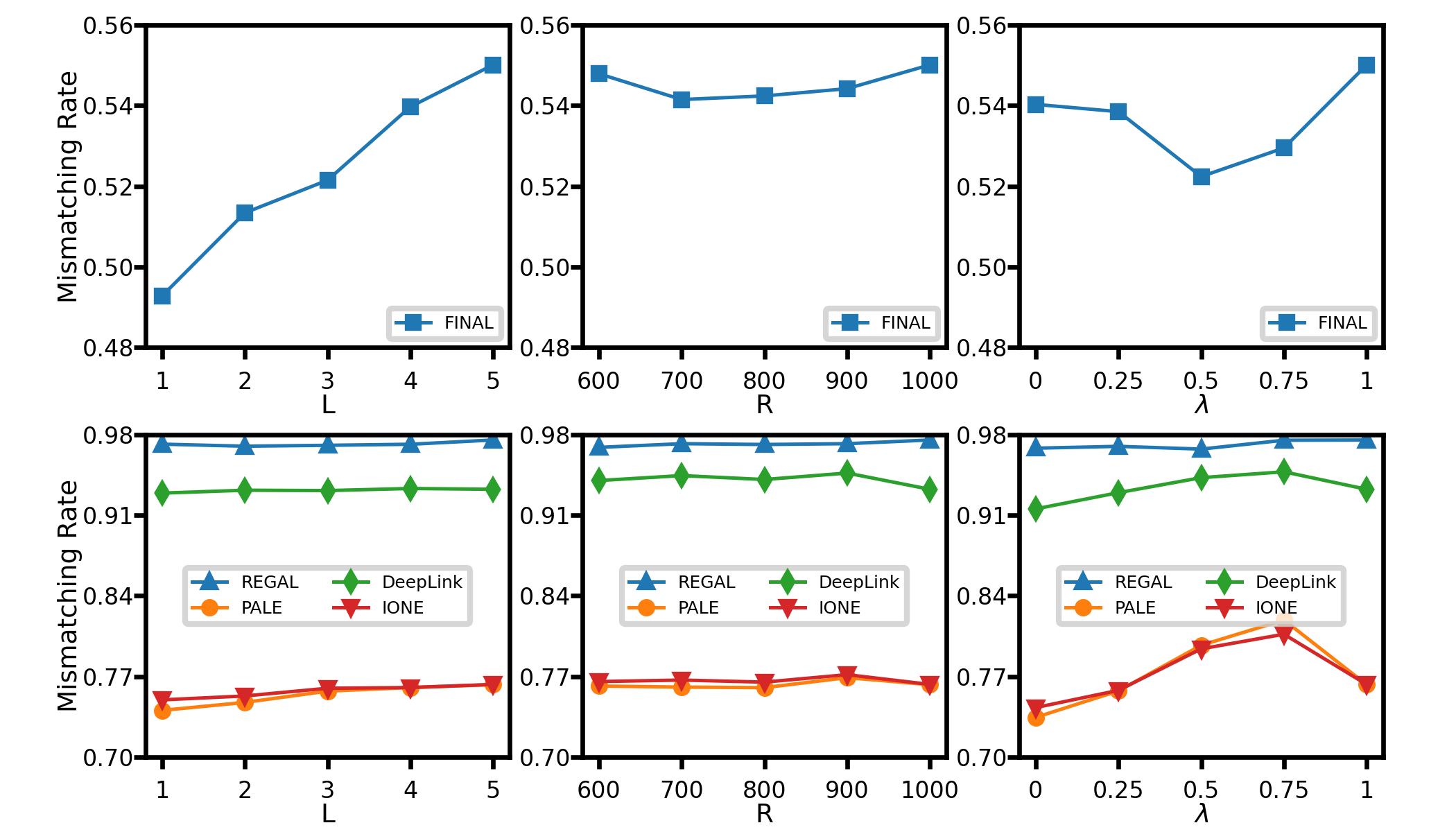}
	\caption{Effect of Hyperparameters on the dataset Douban (online $\rightarrow$ offline)}
	\label{hyperdb}
\end{figure}

\subsection{Effect of hyperparameters on dataset Douban (online $\rightarrow$offline)}\label{appendix_c}

As shown in Fig.~\ref{hyperdb}, we can notice that using a longer walk length can improve the TOAK performance on FINAL, which might be because the longer the length the more features about neighbors are considered, while the performance shows a slight improvement and then becomes stable on other algorithms. In addition, the TOAK model is not sensitive to the random walk number. Finally, the influence of $\lambda$ is also important. On the FINAL and REGAL algorithms, the mismatching rate first decreases and then increases. However, the trend is opposite on PALE, DeepLink, and IONE, where too large and too small $\lambda$ will all worsen the performance.

\end{document}